\documentclass[aps,pre,showpacs,twocolumn,superscriptaddress,showkeys,reprint]{revtex4-1}
\usepackage{amsmath,amssymb,graphicx}
\usepackage[usenames,dvipsnames]{color}
\usepackage[colorlinks=true,linkcolor=Red,citecolor=Green]{hyperref}

\let\Im=\Imag
\DeclareMathOperator{\vol}{vol}

\DeclareMathOperator{\DEC}{DEC}
\DeclareMathOperator{\QNM}{QNM}

\let\Re=\Real

\DeclareMathOperator{\const}{const}

\begin{document}
\title{Trapping of waves and null geodesics for rotating black holes}
\author{S.~Dyatlov}
\affiliation{Department of Mathematics, Massachusetts Institute of
  Technology, Cambridge, MA 02139, USA}
\author{M.~Zworski}
\affiliation{Department of Mathematics, University of California,
  Berkeley, CA 94720, USA}

\date{\today}
\begin{abstract}
We present dynamical properties of linear waves and null geodesics
valid for Kerr and Kerr--de Sitter
black holes and their stationary perturbations. The two are intimately linked
by the geometric optics approximation.
For the nullgeodesic flow the key property is the {\em
  $r$-normal hyperbolicity} of the trapped set and for linear waves
it is the distribution of quasi-normal modes: the exact quantization
conditions do not hold for perturbations but the bounds on 
decay rates and the statistics of frequencies are still valid. 
\end{abstract}


\maketitle

\begin{figure}
\includegraphics[width=7cm]{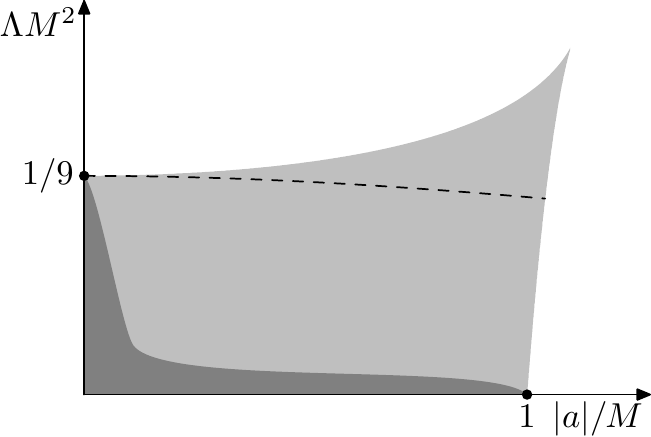}
\caption{Numerically computed admissible range of 
parameters for the subextremal Kerr--de Sitter black hole
(light shaded) \cite{29-a-m} and the range to which our
results apply (dark shaded). QNM are defined and
discrete for parameters below the dashed line, 
$ ( 1 - \alpha )^3 = 9 \Lambda M^2$, see \cite[\S 3.2]{8-Dya11}.}
\label{f:kdsu}
\end{figure}

\addtocounter{section}{1} 
The Kerr solutions \cite{1-Kerr} to Einstein equations 
are considered as physically relevant models of rotating
black holes. The Kerr metrics depend on two parameters:
mass $ M$ and rotational parameter $ a $; the special case $a=0$
is the Schwarzschild metric. The
\emph{Kerr-de Sitter} solutions describe rotating black holes
in the case of positive cosmological constant, $\Lambda>0$~-- see \eqref{eq:Gr}
below for the formula for the metric and Fig.~\ref{f:kdsu} for the
plot of admissible values of the parameters. Due to the observed
{\em cosmic acceleration} \cite{2-Perl}, the current cosmological
$\Lambda$CDM model assumes $ \Lambda > 0 $.
As explained below, $ \Lambda >
0 $ makes the study of the topic of this article,
\emph{quasi-normal modes} for black holes (QNM), mathematically more
tractable while not affecting the description of the physical 
phenomenon of ringdown  \cite{3-SK}.

The classical dynamics of Kerr black holes is concerned with the behavior
of null geodesics of the corresponding metric, that is, the trajectories of photons in 
the gravitation field of the black hole.
The key dynamical object is the {\em trapped set},
consisting of all null geodesics in phase space (position and momentum space)
which never cross the event horizon of the black hole or escape to infinity.
In other words,
this is the set where the strength of gravitational fields forces
photons to travel on bounded orbits.

\begin{figure}
\includegraphics[width=8cm]{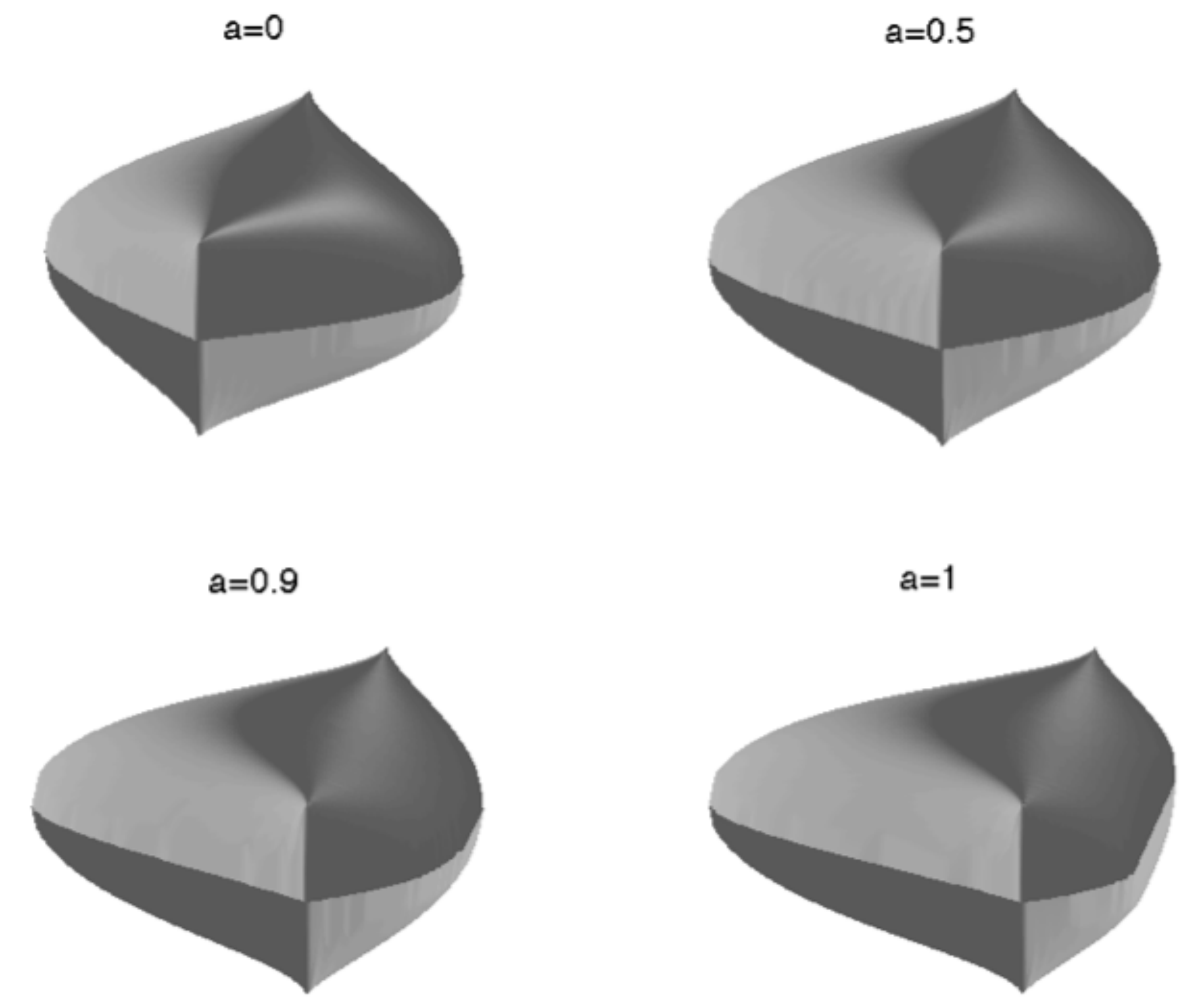}
\caption{Visualization of the trapped set for different
values of $ a $, with $ \Lambda = 0 $. The
figures show the {four-dimensional} set $ K \cap \{ \xi_t = 1 \} $ 
($ K $ is the five dimensional trapped set)
projected to the coordinates
$ ( x, y , z) = ( \xi_\varphi, \theta, \xi_\theta ) $. 
For $ a = 0 $ this corresponds to the visualization of
the phase space of the $2$-sphere: the sphere is  parametrized by 
the coordinates $ 0 \leq  \theta \leq \pi $, $ 0 \leq \varphi < 2
\pi $, $ ( \sin \theta\cos \varphi , \sin \theta \sin \varphi , \cos
\theta ) $. The conjugate coordinates are denoted $ \xi_\theta $ and $
\xi_\varphi $ and the restriction to $ \xi_t  = 1 $ means that
$ \xi_\theta^2 + \sin^{-2} \theta \,\xi_\varphi^2 = { 27M^2
} $.
The vertical
singular interval in front corresponds
to $ \theta = 0 $, with the symmetrical interval in the
back corresponding to  $ \theta = \pi $: 
the coordinates $ ( \theta , \varphi ) $ on the sphere
are singular at that point.  
The structure of $ K $ becomes more interesting when 
$ a > 0 $ as shown in the three examples. 
The additional coordinate,
not shown in the figures, 
$r $ is a function of $ \xi_\varphi $ and $ \xi_t $ only.
When $ a = 0 $, we have $ r = 3 M $, but $r$ gets larger to 
the left ($ \xi_\varphi > 0 $) and smaller to the right 
($ \xi_\varphi < 0 $) when $ a > 0 $. 
When $ a = 1 $ we see the flattening in the 
$ (\theta , \xi_\theta) $-plane at extremal values of
$ \xi_\varphi $: the trapped set touches the event horizon
$ r = 1 $ which results in lack of decay, and some 
 QNM have null imaginary parts \cite{21-hod,22-yc}. Dynamically
and invariantly this corresponds to the vanishing expansion 
rates -- see Fig.~\ref{f:van}.}
\label{f:trapped}
\end{figure}

In case 
of Schwarzschild black hole ($ a = 0 $) the time slice of the trapped set is just the phase
space of a sphere (mathematically, the cotangent bundle of a sphere)
called the {\em photon sphere}: along the photon sphere, all photons travel on closed
orbits. A traveller who crosses
the photon sphere, although still visible to outsiders, is forced to 
cross the black hole horizon eventually. In the case of 
nonzero angular momentum ($ a \neq 0 $) the trapped set
is no longer the phase space/cotangent bundle of a smooth spatial
set; instead it becomes a non-trivial object in the phase space. The photons
are {\em trapped} because of the strength of the gravitational field
but most of them (that is, a set of full measure) no longer travel along closed orbits -- see Fig.~\ref{f:trapped}
for a visualization of the trapped set and \eqref{eq:Ktr} for the 
analytic description. Although the trapped set is no longer the phase
space
of a spatial object, it remains a
smooth five dimensional manifold. The symplectic form on the
phase space of a time slice (see \eqref{eq:sympl}) restricts to a non-degenerate
form on the trapped set. That means that the time slice of the trapped set is a smooth
symplectic manifold.

A remarkable feature of the geodesic flow on
Kerr(--de Sitter) metrics is its complete integrability \cite{4-Carter} in the
sense of Liouville-Arnold \cite{5-Arnold?}: there exist action variables
which define invariant tori on which the motion is linear.

In this article we describe another important 
feature of the dynamics: {\em $r$-normal hyperbolicity}.
It means that the flow is hyperbolic in 
directions normal to the trapped set in ways
$r$-fold stronger than the flow on the trapped set -- see
\eqref{e:rnh-1} for a mathematical definition.
This property, unlike complete integrability, is known to be stable
under perturbations~\cite{6-HPS}: a small $ C^r $ ($r$ times
differentiable) 
stationary perturbation of the
 metric will destroy complete integrability but will
preserve $ C^r $
structure of the trapped set and $r$-normal hyperbolicity.
For Kerr black holes the condition holds for {\em each} $ r $ and hence
regular perturbations will maintain the regular structure of the trapped
set of light trajectories \cite{7-WuZ,8-Dya11}.

The classical dynamical features are crucial for the behavior
of gravitational waves emitted by black holes, especially during the \emph{ringdown} phase,
when a black hole spacetime settles down after a
large cosmic event such as a binary black hole merger.
Gravitational waves are
expected to be observable by the existing detectors, once they are
running at full capacity, and to provide information about the
parameters of astrophysical black holes.
During the ringdown phase, the behavior of gravitational waves is driven
by the linearized system~\cite{3-SK} and is much simpler to simulate numerically than the
merger phase~\cite{9-BCS,10-Ringdown,11-campanelli}. At ringdown,
gravitational waves have a fixed set of complex frequencies,
known as 
\emph{quasi-normal modes (QNM)} \cite{3-SK}
and depending only on the parameters of the black hole, rather
than the specifics of the event.
The simplest model of ringdown is obtained by solving the linear scalar wave
equation for the Kerr(--de Sitter) black hole spacetimes, and in 
that case quasi-normal modes can be rigorously defined. More
complicated linearizations have also been studied \cite{3-SK,12-BK} but 
we concentrate on the simplest setting here. On the relevant time 
and space length scales, the value of the cosmological constant
$ \Lambda $ does not have a physical effect on the ringdown since
gravitational waves are generated in a neighborhood of the black hole
but $ \Lambda > 0 $ makes the mathematical definition of QNM 
much easier by eliminating the polynomial fall-off for waves~\cite[\S5.1]{3-SK}.

In a more general physical or geometric context of scattering theory,
quasi-normal modes, also known as resonances, replace bound states (eigenvalues), for 
systems which allow escape of energy \cite{13-mz} and simultaneously 
describe oscillations (real parts of the mode) and decay (imaginary 
parts). They appear in expansions of waves -- see \eqref{eq:exp} below,
just as waves in bounded regions are expanded using
eigenvalues. This dynamical interpretation immediately suggests
that the distribution of QNM is related to the trapping on the 
classical level -- see \cite{14-Pot,15-Bark} for a discussion and recent
experimental results in the setting of microwave billiards.

The relation between dynamics and distribution of QNM/resonances has
been particularly well studied in problems where a reduction to one
dimension is possible. More generally, complete integrability 
allows quantization rules which can be used to describe resonances in
the semiclassical/high energy limit. In the setting of black holes
this goes back to \cite{16-IS}.

For Schwarzschild black holes the 
Regge--Wheeler reduction (see \eqref{eq:Schr} below) produces a one dimensional potential similar
to the Eckart barrier potential $ \cosh^{-2} x $ for which resonances are
given by ${ \pm {\sqrt{3}}/2 - i ( n + 1/2 )} $ -- see \cite{17-ZZ} for a 
review in the context of chemistry, \cite{18-SZ} for a mathematical 
discussion in the Schwarzschild case,
and~~\cite{18.5-beyer} for a general study of P\"oschl--Teller potentials. Putting together different
angular momenta produces an (approximate) lattice of resonances.

When $ a \neq 0 $, that is, in the genuine Kerr case, the degenarate QNM split in a way similar to the
Zeeman effect. They have been recently studied using WKB methods based
on the completely integrable structure \cite{19-Dya01,20-Dya02,21-hod,22-yc} and
the Zeeman-like splitting has been rigorously confirmed.

The point of this article is to describe recent mathematical
results \cite{18-SZ,7-WuZ,19-Dya01,20-Dya02,23-Dya1,24-Va,8-Dya11}  which apply to {\em stationary perturbations of Kerr metrics} and do not
depend on  the completely integrable structure. They are based on 
the use of the $r$-normal hyperbolicity of the trapped set and 
show that
many features of QNM studied using WKB methods available
in the completely integrable case persist
for perturbations.
The $ r$-normal hyperbolicity of black hole dynamics \cite{7-WuZ} has not been discussed
in the physics literature but the importance of normal hyperbolicity 
in molecular dynamics has been explored~\cite{25-GSWW10}.
It would be interesting to consider the stability
of $r$-normally hyperbolic dynamics under more general, non-stationary, perturbations.

In particular, 
we show how dynamical features (such as the maximal and minimal 
expansion rates) and statistical properties of the distribution of
quasi-normal modes (QNM) depend on $ a = J/Mc $, the rescaled angular momentum.
For the exact Kerr (or Kerr--de Sitter) black hole the counting law for QNM and their
maximal and minimal decay rates determine the mass and the angular
momentum. For the perturbed case, they determine stable features such as the symplectic
volume of the trapped set and classical decay rates. 

The presentation is organized as follows: in \S \ref{tsn} we discuss
classical dynamics and define $r$-normal hyperbolicity in a precise
way; in \S \ref{ddq} we describe the challenges of rigorously defining
of  QNM for Kerr and Kerr-de Sitter black holes. The difficulties 
come from the presence of the ergosphere which obstruct standard
methods for defining resonances (lack of coercivity/ellipticity of the
stationary wave equation) and from the ``size'' of infinity in the 
Kerr ($ \Lambda = 0 $) case. In \S \ref{dqn} we present the 
quantitative results about the distribution of QNM valid for
perturbations of Kerr black hole: bounds on imaginary parts of
the modes \eqref{eq:boundim},  the counting law \eqref{eq:Weyl}, 
and the consequences for solutions of the wave equation
\eqref{eq:exp2}. The strongest results are subject to a pinching
condition \eqref{eq:pinch} which is valid for all but rapidly rotating
rotating black holes. The results presented here and some
of the figures have appeared in works aimed at the mathematical
audience \cite{8-Dya11} and this is an attempt to relate them 
to an active field of research in physics.

\section{The trapped set of null geodesics}
\label{tsn}


We are interested in null geodesics and the wave equation $ \Box_g u =
0 $, where $ g $ is the Kerr(--de Sitter) metric \cite{1-Kerr,4-Carter,7-WuZ,8-Dya11}.
The Kerr--de Sitter metric is a generalization of Kerr to the case
of a positive cosmological constant $\Lambda$.
For a black hole of 
mass  $M>0$ (for $ \Lambda> 0 $ there is no unique definition of
  global mass; here we treat $ M$ simply as a parameter of the metric), rotating with speed $ a $ the space slice is 
$$ X = ( r_+ , r_C ) \times \mathbb S^2 , $$
 where $ r_C < \infty $ when
$\Lambda>0$, and $ r_C = \infty
$ when $ \Lambda = 0 $.  
The behavior of $ \Box_g $ for $ r $ near $
r_C $ is dramatically different in the two cases: for $ \Lambda > 0 $ the
metric is asymptotically hyperbolic in the sense of non-Euclidean geometry (infinity is {\em large} in the sense
that the volume of balls grows exponentially in radius) and for $ \Lambda = 0 $
it is asymptotically Euclidean (infinity is {\em small} in the sense that volume of balls
grows polynomially). For solutions to the wave equation that 
produces differences in long time decay and in the behavior at low
energies \cite{26-Pri,27-Tat,28-Dya2}.
The surface $ r = r_+ $ is the event horizon 
of the black hole. When $ \Lambda > 0 $, $ r = r_C $ is the
cosmological horizon. While the two horizons have different physical
interpretations, their mathematical roles in the study of wave decay
and QNM are remarkably similar.

The geodesic flow can be considered as a flow  on the phase space (the
position-momentum space) of $ \mathbb R 
\times X $, that in mathematical terms on the 
cotangent bundle $T^*(\mathbb R\times X)$.
We denote the coordinates on $ \mathbb R \times X $ by $ ( t,r, \theta, \varphi ) $ (see
Fig.~\ref{f:trapped}), and write $ (\xi_t, \xi_r , \xi_\theta, \xi_\varphi )
$ for the corresponding conjugate (momentum) variables.
The flow is given by the classical Hamiltonial flow \cite{5-Arnold?}
for the
Hamiltonian $ G $,
\begin{gather*}  \dot t = \partial_{\xi_t } G , \ \  \dot r = \partial_{\xi_r} G, \ \  \dot \theta = \partial_{\xi_\theta}
G, \ \ \dot \varphi = \partial_{ \xi_\varphi} G , \\ \dot \xi_t =
- \partial_t G ,  \ \ \dot \xi_r =
- \partial_r G , \ \ \dot \xi_\theta = - \partial_\theta G , \ \ 
\dot \xi_\varphi = -\partial_\varphi G , \end{gather*}
where
\begin{gather}
\label{eq:Gr}
\begin{gathered}
G=\rho^{-2}(G_r+G_\theta), \ \ \rho^2=r^2+a^2\cos^2\theta, \\
G_r=\Delta_r\xi_r^2-{(1+\alpha)^2\over\Delta_r}((r^2+a^2)\xi_t+a\xi_\varphi)^2,\\
G_\theta=\Delta_\theta\xi_\theta^2+{(1+\alpha)^2\over\Delta_\theta\sin^2\theta}(a\sin^2\theta\,\xi_t+\xi_\varphi)^2,
\ \ 
\alpha={\Lambda a^2\over 3}.
\\
\Delta_r=(r^2+a^2)\Big(1-{\Lambda r^2\over 3}\Big)-2Mr,\quad
\Delta_\theta=1+\alpha\cos^2\theta.
\end{gathered}
\end{gather}
The function $G$ is the  dual metric to the semi-Riemannian Kerr(-de
Sitter) metric $g$ (see for example~\cite[\S3.1]{8-Dya11} for
the formulas for $g$).
It is also the principal symbol
of $\Box_g$ in the sense that 
$$\Box_g=G(t,r,\theta,\varphi,\partial_t/i,\partial_r/i,\partial_\theta/i,\partial_\varphi/i),  \ \ i = \sqrt{ - 1} , $$
modulo a first order differential operator. The limiting radii $r_+,r_C$ solve
$\Delta_r=0$.

The trapped set consists of null geodesics that stay away from $r=r_+,r=r_C$
for all times. 
The variables $(\theta,\xi_\theta)$ evolve according to the Hamiltonian flow
of $G_\theta$,  and $\xi_t,\xi_\varphi$ are conserved.
The trapping depends on the evolution of $(r,\xi_r)$
according to the flow of $G_r$, which is essentially the one
dimensional motion for a barrier top potential \cite{18-SZ}.
Under the assumptions that either $a=0$, $ 9\Lambda M^2<1$
or $\Lambda=0$, $|a|<M$, and for nearby values of $M,\Lambda,a$
\cite[Prop.3.2]{8-Dya11},\cite{29-a-m} the
trapped set $K$ is given by 
\begin{equation}
\label{eq:Ktr}
K=\{G=0, \ \xi_r=0,\ \partial_r G_r=0,  \ \xi\neq 0\}.
\end{equation}
For $a=0$, $ \partial_r G_r = 0 $ gives $r=3M$, the radius of the
photon sphere. For $a\neq 0$, a more careful analysis is required, but $K$ is still
a smooth submanifold of the charateristic set $\{G=0\}$. Moreover, it is symplectic
in the sense that the spatial symplectic form $ \sigma $, 
\begin{equation}
\label{eq:sympl} \sigma = d \xi_r\wedge dr+d\xi_\theta\wedge
d\theta+d\xi_\varphi\wedge d\varphi, \end{equation}
is nondegenerate on the surfaces $K\cap\{t=\const\}$.

\begin{figure}
\includegraphics{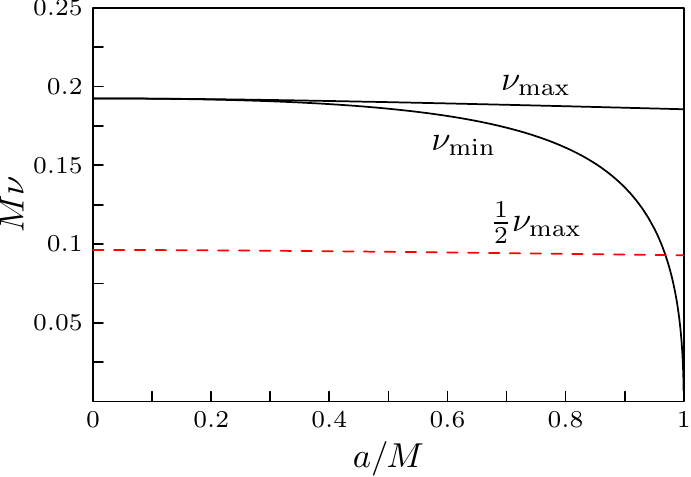}
\caption{The dependence of $ \nu_{\max} $ and $ \nu_{\min} $ on the
  parameters $ M $ and $ a $ in the case of $ \Lambda = 0 $.  The
  dashed line indicates the range of validity of the pinching condition needed for the
  Weyl law \eqref{eq:Weyl}.}
\label{f:1}
\end{figure}

Let $\mathcal C_+\subset \{G=0\}$ be the positive light cone
and 
$$\varphi^t:\mathcal C_+\to\mathcal C_+ $$ 
 the geodesic flow parametrized by $t$.
The $r$-normal hyperbolicity condition asserts the existence of a
$C^r$ ($r$-times differentiable) splitting
\[
T_K\mathcal C_+=TK\oplus \mathcal V_+\oplus\mathcal V_-,
\]
invariant under the flow and such that for some constants $\nu>0,C>0$,
\begin{gather}
\label{e:rnh-1}
\begin{gathered}
\sup_{(x,\xi)\in K}
|d\varphi^{\mp t}
|_{\mathcal V_\pm}|\leq Ce^{-\nu t},\ 
\\
\sup_{(x,\xi)\in K}
|d\varphi^{\pm t}
|_{TK}|\leq Ce^{\nu|t|/r},\ 
t\geq 0. 
\end{gathered}
\end{gather}
This means that the maximal expansion rates (Lyapunov exponents) on
the trapped set are $r$-fold dominated by the expansion and
contraction rates in the directions transversal to the trapped set. 
As shown in \cite{6-HPS},\cite[\S 5.2]{23-Dya1}  $r$-normal
hyperbolicity is stable under perturbations: when $  G_\epsilon $ is a
time independent (that is, stationary) Hamiltonian such that $ G_\epsilon  $ is close to $ G $
in $ C^r $ near $ K $, the flow for $ G_\epsilon $ is $
r$-normally  hyperbolic in the sense that the trapped set $ K_\epsilon
$ has $ C^r $ regularity, is symplectic and \eqref{e:rnh-1} holds.
For Kerr(--de Sitter) metrics the flow is $r$-normally 
hyperbolic for all $ r $ as shown in \cite{7-WuZ,24-Va,8-Dya11}, essentially 
because the flow on $K$ is completely integrable. 

\begin{figure}
\includegraphics[width=8cm]{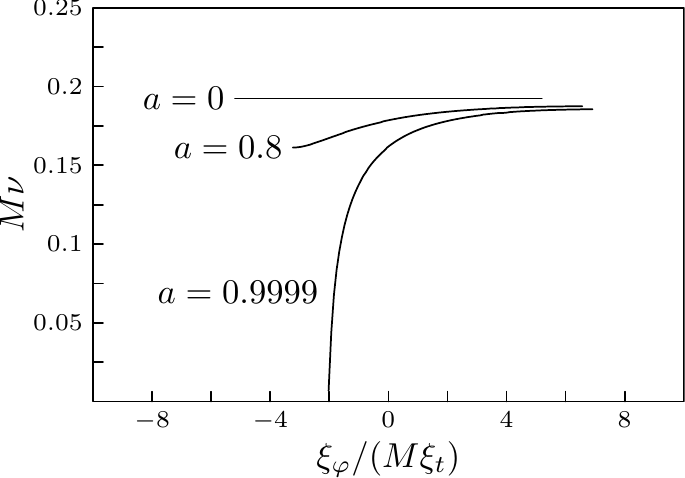}
\caption{The pointwise
expansion rates $ \nu $ on Liouville tori $ \xi_\varphi/
 ( M \xi_t )  = \const $, for $\theta=\pi/2$. When $ a $ approaches
$ 1 $, $ \nu = 0 $ for some values of $ \xi_\varphi $
which shows that there is no gap and QNM can be arbitrarily 
close the real axis \cite{21-hod,22-yc}.}
\label{f:van}
\end{figure}

Key dynamical quantities
are the minimal and maximal expansion rates $0<\nu_{\min}\leq\nu_{\max}$, characterized by inequalities
true for all $\varepsilon>0$, a constant $C_\varepsilon$ depending on $\varepsilon$,
\[ 
C_\varepsilon^{-1} e^{-(\nu_{\max}+\varepsilon)t}\leq
|d\varphi^t|_{\mathcal V_-}|\leq C_\varepsilon
e^{-(\nu_{\min}-\varepsilon)t} , \]
$ t>0 $.
For Kerr(--de Sitter) metrics, the quantities $\nu_{\min},\nu_{\max}$
are obtained by taking the minimum and maximum of averages of the local expansion rate
\[ \nu={\sqrt{-2\Delta_r \partial_r^2G}/|\partial_{\xi_t}G|},  \]
on the Liouville tori \cite{5-Arnold?} of the flow of $G_\theta$ on the trapped set.

\section{Definition and discreteness of quasi-normal modes}
\label{ddq}

The \emph{scattering resonances}, called \emph{quasi-normal modes} (QNM) in
the context of black holes \cite{12-BK}, replace eigenmodes when one
switches from closed systems to open systems -- see \cite{14-Pot} for a
recent experimental discussion. 
They are the frequencies $\omega$ of oscillating solutions to the wave equation
\begin{equation}
  \label{e:equla}
\Box_g (e^{-i\omega t} v(r,\theta,\varphi))=0,
\end{equation}
which continue smoothly across the event horizons. 

Solutions to $\Box_g u =
0 $ are expected to have expansions 
\begin{equation}
\label{eq:exp}   u ( t, r,\theta,\varphi ) \sim \sum_k e^{ - it \omega_k }
 u_k ( r,\theta,\varphi ) \end{equation}  valid in a suitable sense \cite{30-BH,20-Dya02}. 
The fact that QNM $ \omega_k $ form a  discrete set in the lower
half-plane
 is nontrivial but it is now
rigorously known in the case of Kerr--de Sitter and its  perturbations
\cite{31-Bach,19-Dya01,18-SZ,30-BH,24-Va}. 

In the simpler Schwarzschild--de Sitter case 
we indicate the reason for discreteness of the set of QNM as follows.
The equation~\eqref{e:equla} can be rewritten as $P(\omega)v=0$, where
$P(\omega)$ is obtained from $-\rho^2\Box_g$ by replacing $\partial_t$ with $-i\omega$.
The operator $P(\omega)$ is spherically symmetric; its restriction to the space or spherical
harmonics with eigenvalue $\ell(\ell+1)$, written in
the Regge--Wheeler coordinate $x$~\cite[\S4]{19-Dya01}, is
the Schr\"odinger operator
\begin{equation}
\label{eq:Schr}
P_{\ell}(\omega)=-\partial_x^2+\omega^2V_1(x)+\ell(\ell+1)V_2(x) , \end{equation}
where the potentials $ V_1 $ and $ V_2 $ are real analytic (their
Taylor series converge to their values) and satisfy
\[ V_1(x)=-V_\pm^2+\mathcal O(e^{-A_\pm|x|}), \ \  V_2(x)=\mathcal
O(e^{-A_\pm|x|}), \]
as $  x\to \pm\infty$; 
here $ A_\pm>0$, $V_-=r_-^2$, $V_+=r_C^2$.
(When $ \Lambda = 0 $ then $V_2 \sim x^{-2}$ as $x\to +\infty$ and that
creates problems at low energies \cite{26-Pri,27-Tat}. More precisely, it is expected that due
to the slow decay of the potential, the resolvent
is not holomorphic in a neighborhood of zero and even in simplest cases such as Schwarzschild, there is no
mathematical argument excluding the possibility of accumulation of QNM at zero.)
A number $\omega\in \mathbb C$ is a QNM if there exists
an angular momentum
$\ell$ and a nonzero solution $v(x)$ to the equation $P_\ell(\omega) v=0$
 satisfying the
outgoing condition:
near $x=\pm\infty$,
$ e^{\mp iV_\pm \omega x} v(x) $ is a smooth function of  $e^{-A_\pm|x|}$.
The outgoing condition follows naturally from the requirement that $e^{-i\omega t}v(x)$ extends
smoothly past the event horizon of the black hole.
For fixed $\ell$, it follows by standard one-dimensional methods that the set of all corresponding
$\omega$ is discrete.

Showing that as $\ell\to\infty$,
quasi-normal modes
corresponding to different values of 
 $\ell$ do not accumulate is more delicate:
we need to know that if $|\omega|\leq R$ and $\ell$  is large enough depending on $R$,
then $\omega$ cannot be a QNM corresponding to $\ell$. 
Assume the contrary and let $v(x)$ be the corresponding solution to the
equation $P_{\ell}(\omega)v=0$. We fix large $X>0$ independently of $\ell$, to be chosen
later. The potential $V_2$ is everywhere positive, therefore for $\ell$
large enough depending on $R,X$,
$ \Re (\omega^2V_1(x)+\ell(\ell+1)V_2(x))>0\quad\text{for }x\in
[-X,X]. $
If $v$ satisfied a Dirichlet or Neumann boundary condition at $\pm X$,
then integration by parts would give the impossible statement that
$v=0$ on $[-X,X]$, finishing the proof:
$$
\begin{gathered}
0=\Re\int_{-X}^X \overline{v}\cdot P_\ell(\omega) v \,dx
=-\Re(v'\bar{v})\big|_{-X}^X\\
+\int_{-X}^X |v'|^2+\Re (\omega^2V_1+\ell(\ell+1)V_2)|v|^2\,dx=0,
\end{gathered}
$$
and the terms under the integral are all nonnegative. 
This argument works also for $ v $'s satisfying the
defining properties of the QNM, as described below. For Kerr--de Sitter
black holes a separation procedure is still possible but 
it does not work 
for stationary perturbations.
Nevertheless in both cases
the discreteness
of QNM remains valid \cite{23-Dya1,24-Va}. 

To indicate how this works for resonant states which not satisfy a boundary condition at $\pm X$ (after all, this
`boundary' is completely artificial), we follow \cite[\S6]{19-Dya01}.
It suffices
to prove the boundary inequalities
\begin{equation}
  \label{e:bacon}
\pm\Re(v'(\pm X)\overline{v(\pm X)})<0.
\end{equation}
To prove~\eqref{e:bacon}, we cannot use integration by parts on the whole $\mathbb R$,
since $v$ does not lie in $L^2(\mathbb R)$ and moreover the real part of our
potential may become negative as $x\to\pm\infty$. We instead use the methods of complex
analysis and real analyticity of $V_1,V_2$.

The characterization of $ v$ as a mode can be strengthened to
say that $e^{\mp iV_\pm\omega x}v(x)$ is an \emph{real analytic}
function of $e^{-A_\pm|x|}$ (it has a convergent Taylor series in that
variable),
which means that for $X$ large enough, we can
extend $v(x)$ to a \emph{holomorphic} function in $\{|\Re z|\geq X\}$, and this
extension is Floquet periodic:
$ v(z+2\pi i/A_\pm)=e^{\mp 2\pi V_\pm\omega/A_\pm}v(z)$, 
$ \pm \Re z\geq X$.
Now, consider the restriction of $u$ to the vertical lines $\{\Re z=\pm X\}$,
$w_\pm(y):=v(\pm X+iy)$, $y\in\mathbb R$,
and note that it solves the differential equation
\begin{equation}
  \label{e:wildcat}
(\partial_y^2+\omega^2V_1(\pm X+iy)+\ell(\ell+1)V_2(\pm X+iy))w_\pm =0.
\end{equation}
The key difference between~\eqref{e:wildcat} and the equation $P_{\ell}(\omega)v=0$ is
that the potential $V_2(\pm X+iy)$ is no longer real-valued. For instance,
if $V_2$ were equal to $e^{-A_\pm|x|}$, then
$V_2(\pm X+iy)$ would equal $e^{-XA_\pm}e^{\mp iA_\pm y}$, only
taking real values when $y\in \pi A_\pm^{-1}\mathbb Z$. This means
that the equation~\eqref{e:wildcat} is \emph{elliptic} (in the semiclassical
sense, where we treat $\partial_y$ as having same order as $\ell$) except at
a discrete set of points in the phase space $T^*\mathbb R$. Further analysis shows
that $w_\pm(y)$ is concentrated in phase space near
$y\in 2\pi A_\pm^{-1}\mathbb Z,\eta=\mp \ell \sqrt{V_2(\pm X)}$, in particular implying
$$
|(\partial_y \pm i\ell\sqrt{V_2(\pm X)})w_\pm(0)|\leq C\ell^{1/4}|w_\pm(0)|
,$$
and~\eqref{e:bacon} follows from Cauchy--Riemann equations, since
$v(\pm X)=w_\pm(0)$ and $v'(\pm X)=-iw_\pm'(0)$.

\section{Distribution of quasi-normal modes}
\label{dqn}

The distribution of QNM $ \omega_k $ can now be
studied in the more general stable setting of $r$-normally hyperbolic
trapped sets. Three fundamental issues are:

\noindent
{\em (a) } distribution of
decay rates, that is of the imaginary parts of QNM;

\noindent
 {\em (b)}
asymptotics of the counting function;

\noindent {\em (c)}  expansion of waves in
terms of QNM. 

\begin{figure}
{\vspace{0.2in} \includegraphics[width=7cm]{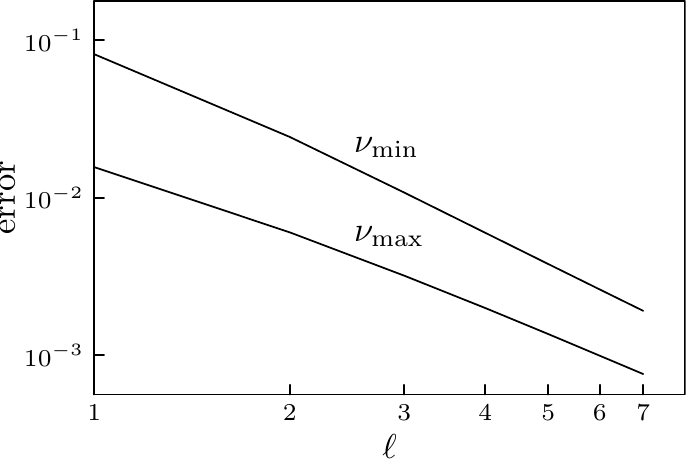}}
\caption{A log-log plot of relative errors 
$ | \min  | \Im \omega_k (  \ell ) | - \nu_{\min}/2 |\over \min | \Im \omega_k (
    \ell ) |$  and 
$
  | \max  | \Im \omega_k (
    \ell ) | - \nu_{\max}/2 |\over \max | \Im \omega_k (
    \ell ) |$   where $ \omega_k ( \ell ) $ are the numerically computed
  resonances in the first band corresponding to the angular momentum 
$ \ell $ \cite{9-BCS} and $ \nu_{\min} $,$ \nu_{\max} $ are minimal and
maximal expansion rates. The agreement is remarkable when $ \ell $ increases.}
\label{f:2}
\end{figure}

\begin{figure}
\includegraphics[width=7cm]{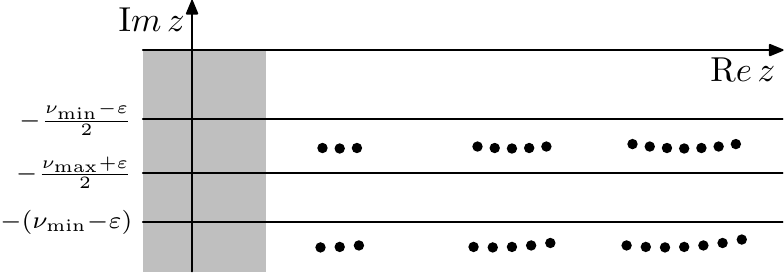}
\vskip.1in
\includegraphics[width=7cm]{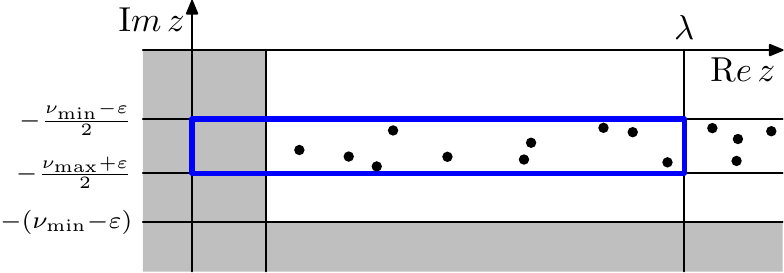}
\caption{A schematic comparison between QNM in a
  completely integrable case and in the general $r$-normally hyperbolic
  case. The former lie on a fuzzy lattice and are well approximated
by WKB construction based on quantization conditions
\cite{19-Dya01,20-Dya02,21-hod,22-yc}. When trapping is $r$-normally hyperbolic and
the pinching condition $ \nu_{\max} < 2 \nu_{\min} $ holds,
quasi-normal modes are still localized to a strip with dynamically 
determined bounds, and their statistics are given by the Weyl law \eqref{eq:Weyl}.}
\label{f:iv}
\end{figure}

\medskip
\noindent
{\em (a)} We can bound the decay rates from below whenever the trapped
set is normally hyperbolic, without requiring the stronger $r$-normal 
hyperbolicity assumption. The bound~\cite{32-NZ2} is given by $ \Im
\omega_k < - (\nu_{\min}-\varepsilon)/2 $, for any $ \varepsilon > 0
$, once the frequency (the real part of $ \omega_k $) is large enough.
In the case of $ r$-normal hyperbolicity 
and under the pinching condition 
\begin{equation}
\label{eq:pinch}  \nu_{\max } < 2 \nu_{\min } , 
\end{equation}
we get more detailed information \cite{23-Dya1}: 
there are additionally no
QNM with
\begin{equation}
\label{eq:boundim}  - (\nu_{\min}-\varepsilon) < \Im \omega_k <
-(\nu_{\max}+\varepsilon)/2 .
\end{equation}
That means that the modes with least decay are confined to a band
shown in Fig.\ref{f:iv}. In the completely integrable case this 
follows from WKB constructions \cite{19-Dya01,20-Dya02,21-hod,22-yc} -- see
Fig.\ref{f:iv}~-- but this structure persists under perturbations.
Fig.\ref{f:2} shows the accuracy of the estimate \eqref{eq:boundim} 
for the numerically computed QNM of exact Kerr black holes \cite{9-BCS}.
For a recent experimental investigation of the distribution of
decay rates and the relation to classical dynamics (more precisely,
the topological pressure and classical escape rates) see \cite{15-Bark}.

The condition~\eqref{eq:pinch} is called pinching because it pinches the ratio
of the maximal and the minimal transversal expansion rates. In the absense of this condition,
the gap~\eqref{eq:boundim} between the first band of QNM and the faster decaying bands disappears, making it difficult
to obtain a counting asymptotics~\eqref{eq:Weyl}. Physically, \eqref{eq:pinch} could interpreted as the requirement
that there be no interaction between QNM from different bands.

\medskip
\noindent
{\em (b)} The relation between the density of high energy states and 
phase space volumes defined by the classical Hamiltonian is one of the
basic principles of quantum mechanics/spectral theory. It
states that for closed systems the number, $ N_{\widehat H} ( \lambda ) $, of energy levels of $ \widehat H $, a
quantization of $ H $ (for instance the Dirichlet Laplacian on a
bounded domain), below energy $ \lambda^2 $ (we think of $ \lambda $
as frequency which is natural when considering QNM) satisfies the {\em Weyl law}
\[
N_{ \widehat H} ( \lambda ) \sim ( 2 \pi )^{-\dim X  } \vol_{T^* X} (
H \leq \lambda^2 ) \sim C_{\widehat H} \lambda^{-\dim X }.
\]
Here $ \vol_{ T^*X } $ denotes the phase space volume calculated  using
the volume $ \sigma^{\dim X}/ (\dim X)! $
obtained from the symplectic form $ \sigma $ (see \eqref{eq:sympl}).

For open systems QNM replace real energy levels and the
counting becomes much more tricky \cite{14-Pot}. In the case of exact
 Kerr(--de Sitter) black holes the WKB constructions can be used
to show that
the number, $ N_{\QNM} ( \lambda ) $,  of QNM with 
\[  | \omega_k|  \leq \lambda , \ \  \Im \omega_k \geq
-(\nu_{\min}-\varepsilon) \]  satisfies the asymptotic law
$ N_{\QNM} ( \lambda) \sim c \lambda^{2} $. The constant $ c $ has
a geometric interpretation: in a scattering problem the total phase space $
T^*X $ (the cotangent bundle of $ X $) is replaced by the trapped set
\cite{33-Sj}, and $c$ corresponds to the symplectic volume of the trapped set.

\begin{figure}
\includegraphics[width=8cm]{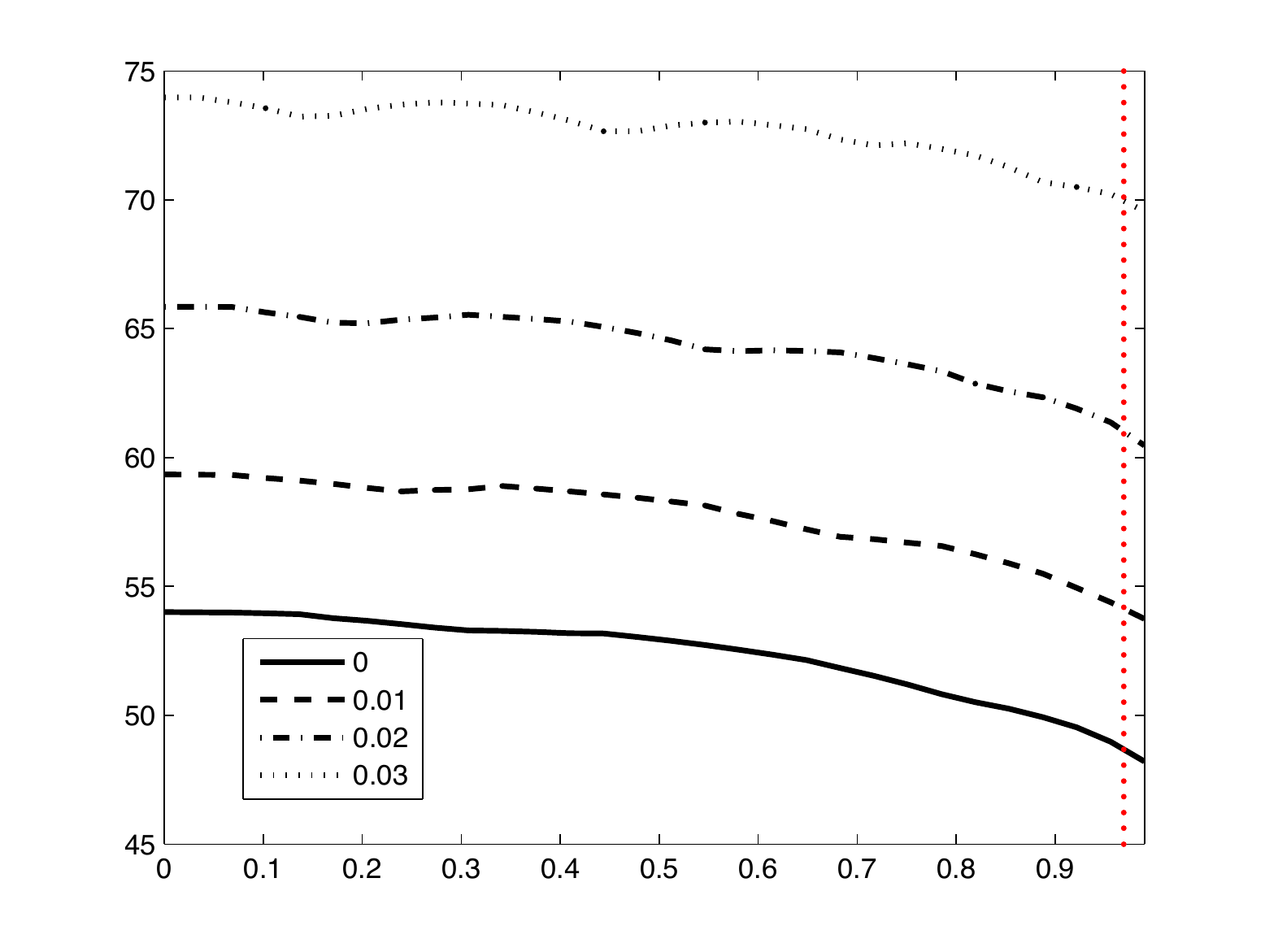}
\caption{Numerically computed constant 
in \eqref{eq:Weyl}, $ \vol(K \cap \{\xi_t^2 \leq 1 \} \cap \{ t = 0 \})/4 \pi^2 $, for $ \Lambda = 0,0.01,0.02,0.03 $ and $ M = 1$. 
The vertical line shows the value of $ a $ at which the pinching
condition~\eqref{eq:pinch} fails for $ \Lambda = 0 $. For smaller values of $ a $ 
the Weyl law~\eqref{eq:Weyl} holds. We note that as $ a$
increases the QNM split (see Fig.~\ref{f:iv}) 
and hence we do expect the counting function to decrease, in agreement with the behavior of the volume. From the volume and the gap giving
the decay rate one can read off $ a$ and $ M$.}
\label{f:vol}
\end{figure}

The same law is proved~\cite{8-Dya11} for perturbations of Kerr--de Sitter,
using completely different ideas based on $r$-normal hyperbolicity rather than
symmetries of the metric and separation of variables.
Under the assumptions of
$r$-normal hyperbolicity \eqref{e:rnh-1} and pinching~\eqref{eq:pinch},
we have
\begin{equation}
  \label{eq:Weyl}
N_{\QNM} ( \lambda ) \sim  {\lambda^{2}\over (2\pi)^2} \vol( K \cap \{ 
\xi_t^2 \leq 1 \} \cap \{ t = 0 \} ), 
\end{equation}
where the volume is taken using the symplectic form on $ K \cap \{ t =
\const \} $ \cite[Thm 3]{8-Dya11}. We note that just as $ \dim X =
\frac12 \dim T^* X $ in the exponent of the Weyl law, here 
\[  2 = \frac12 \dim ( K \cap \{ t = \const \} ) , \]
that is,  the effective phase 
space is now the trapped set. 
For 
exact Kerr(--de Sitter) metrics with several 
values of $ \Lambda $ the volume as function of $ a $ is shown
in Fig.~\ref{f:vol}. The volume is finite provided that
$ ( 1 - \Lambda a^2/3)^3 > 9 \Lambda M^2$, see Fig.~\ref{f:kdsu}.

We should
stress that normally hyperbolic behavior
(unlike $r$-normal hyperbolicity) is often unstable under perturbations
as shown by examples of hyperbolic quotients where a small
perturbation can change the dimension of the trapped set
\cite[Fig.1]{34-DaDy}, leading 
to a fractal Weyl law, that is a law in which the exponent $ 2 $
changes
to half of the fractal dimension of the trapped set  -- see \cite{14-Pot} for recent experiments
on that.

\medskip
\noindent
{\em(c)}  The expansion \eqref{eq:exp} is rigorously established
for slowly rotating black holes \cite{30-BH,20-Dya02} and heuristically 
it is one of the motivations for studying quasi-normal modes
\cite{12-BK}. For rapidly rotating black holes, or for their perturbations
satisfying \eqref{eq:pinch}
a more robust version can be formulated using projector onto
the states associated to quasi-normal modes in the first band shown in 
Fig.~\ref{f:iv}. The solution to the wave equation $ \Box_g u = 0 $
with initial data localized near frequency $ \lambda\gg 1 $ can be
decomposed as $$ u = u_{\QNM} + u_{\DEC} $$ where, for $ 0 \leq t \leq
T \log \lambda$, 
$$ \Box_{ g}u_{\QNM} (t), \ \ \Box_{ g}u_{\DEC} (t) = {\mathcal O}
( \lambda^{-\infty}),  $$ 
that is we have rapid decay (faster than any negative power) when the
frequency $ \lambda $ is large.
This means that both terms solve the wave equation
approximately at high energies times bounded logarithmically in $\lambda$. We then have,
again for  $ 0 \leq t\leq T\log \lambda $, 
\begin{equation}
\label{eq:exp2}  \begin{split} 
& \|u_{\QNM } (t)\|_{\mathcal E} \leq C
e^{-(\nu_{\min}-\varepsilon)t/2} 
\|u_{\QNM} (0)\|_{\mathcal E}, \\
& \|u_{\QNM} (t)\|_{\mathcal E}\geq
C^{-1}e^{-(\nu_{\max}+\varepsilon)t/2}\|u_{\QNM} ( 0 )\|_{\mathcal E}, \\
&\|u_{\QNM} (0)\|_{\mathcal E}\leq C \sqrt\lambda \|u(0)\|_{\mathcal E},\\
& \|u_{\DEC} (t)\|_{\mathcal E}\leq C \lambda
e^{-(\nu_{\min}-\varepsilon)t}\|u(0)\|_{\mathcal E},
\end{split} \end{equation}
where strictly speaking errors $  {\mathcal O} ( \lambda^{-\infty }
)\|u(0)\|_{\mathcal E} $ should be added to the right hand sides. The
norm  $ \| \cdot \|_{\mathcal E}$ is the standard energy norm in any 
sufficiently large compact subset of $ X $ (in the case of exact
Kerr--de Sitter, we can take $ ( r_+ + \delta, r_C - \delta ) \times
\mathbb S^2 $) -- see \cite[Thm 2]{8-Dya11}. The term $ u_{\QNM} ( t ) $
corresponds to the part of the solution dominated by the QNM
in the first band and it has the natural decay properties
dictated by the imaginary parts of these QNM.
In fact, $u_{\QNM}(t)$ can be physically interpreted as the radiation coming from
light rays traveling along the trapped set. The directions in which such a light ray radiates
towards infinity can be described in terms of the geometry of the flow,
and the amplitude of the radiated waves can be calculated using the global dynamics
of the flow near the trapped set~\cite[\S8.5]{23-Dya1}.

\section{Conclusions}
\label{con}
We have shown that for
Kerr--de Sitter metrics and their perturbations quasi-normal modes
are rigorously defined and form a discrete set in the lower half
plane, provided 
that the parameters of the black hole satisfy
\[ ( 1 - \Lambda a^2 / 3)^3 > 9 \Lambda M^2 > 0,\]
see Fig.~\ref{f:kdsu}. 
This is due to the size of infinity when $ \Lambda > 0 $ and
the compactness of the trapped set at finite energies.

If one neglects the issues of long time decay and of behavior
at low energies, then the results are also valid in the case of $
\Lambda = 0 $. On the length scales involved in the ringdown
phenomenon, which in principle would lead to the detection of
black hole parameters through QNM, the (small) value of $ \Lambda $ 
is not relevant but $ \Lambda > 0 $ is a more convenient mathematical
model.

The main dynamical feature of the set on which photons are trapped 
(the trapped set) is its $ r$-normal hyperbolicity for any $ r $ --
see \eqref{e:rnh-1}. Because of the stability of this property
the main features of the distribution of quasinormal modes are preserved
for perturbations: the decay rates are bounded from below in 
terms of the minimal expansion rate ($ \Im \omega_k \leq -
(\nu_{\min}-\varepsilon)/2$) and under the pinching conditions, the least
decaying modes are confined to a strip where they satisfy 
a counting law \eqref{eq:Weyl} -- see Fig.~\ref{f:iv}.

The $ r$-normal hyperbolicity is valid for all rotating black holes
but the pinching condition \eqref{eq:pinch} needed for the finer
results \eqref{eq:Weyl} and \eqref{eq:exp2} fails in the 
case of very fast rotation~-- see Fig.~\ref{f:1}.

\medskip

\noindent
{\sc Acknowledgments.} This work was partially supported by the National Science 
Foundation grant DMS-1201417 (SD,MZ) and by the Clay Research Fellowship (SD). We are grateful to 
Mihalis Dafermos for stimulating discussions of black hole physics and
to St\'ephane Nonnenmacher for comments on earlier versions of this note.
We are also thankful to three anonymous referees for suggesting many improvements in the presentation.

\end{document}